\documentclass[
superscriptaddress,
 amsmath,amssymb,
 aps,
]{revtex4-2}

\usepackage{graphicx}
\usepackage{dcolumn}
\usepackage{bm}
\usepackage[caption=false]{subfig}

\begin{document}

\preprint{APS/123-QED}

\title{Enhanced Yield from a Cryogenic Buffer Gas Beam Source via Excited State Chemistry}

\author{Arian Jadbabaie}
\affiliation{Division of Physics, Mathematics, and Astronomy, California Institute of Technology, Pasadena, California 91125, USA}
\author{Nickolas H. Pilgram}
\affiliation{Division of Physics, Mathematics, and Astronomy, California Institute of Technology, Pasadena, California 91125, USA}
\author{Jacek K{\l}os}
\affiliation{Department of Physics, Temple University, Philadelphia, Pennsylvania 19122, USA}
\affiliation{Department of Chemistry and Biochemistry, University of Maryland, College Park, Maryland 20742, USA}
\author{Svetlana Kotochigova}
\affiliation{Department of Physics, Temple University, Philadelphia, Pennsylvania 19122, USA}
\author{Nicholas R. Hutzler}%
\affiliation{Division of Physics, Mathematics, and Astronomy, California Institute of Technology, Pasadena, California 91125, USA}

\date{\today}

\begin{abstract}
We use narrow-band laser excitation of Yb to substantially enhance the brightness of a cold beam of YbOH, a polyatomic molecule with high sensitivity to physics beyond the Standard Model (BSM). By exciting atomic Yb to the metastable $^3$P$_1$ state in a cryogenic environment, we significantly increase the chemical reaction cross-section for collisions of Yb with reactants. We characterize the dependence of the enhancement on the properties of the laser light, and study the final state distribution of the YbOH products. The resulting bright, cold YbOH beam can be used to increase the statistical sensitivity in searches for new physics utilizing YbOH, such as electron electric dipole moment (eEDM) and nuclear magnetic quadrupole moment (NMQM) experiments. We also perform new quantum chemical calculations that confirm the enhanced reactivity observed in our experiment. Additionally, our calculations compare reaction pathways of Yb($^3$P) with the reactants H$_2$O and H$_2$O$_2$. More generally, our work presents a broad approach for improving experiments that use cryogenic molecular beams for laser cooling and precision measurement searches of BSM physics.

\end{abstract}

\maketitle

\section{Introduction}
Cold, gas-phase molecules represent a rapidly growing resource for the next generation of experiments in atomic, molecular, and optical physics. For example, tabletop experiments utilizing molecules cooled to sub-meV energy scales have successfully constrained symmetry-violating physics beyond the Standard Model at the TeV-scale~\cite{ACME2018,Cairncross2017}, with the possibility on the horizon of reaching PeV-scale limits~\cite{Kozyryev2017PolyEDM}.  Compared to atoms, molecules provide significant advantages for powerful experiments in quantum simulation, quantum information, quantum many-body systems, and searches for physics beyond the Standard Model (BSM)~\cite{Carr2009,Quemener2012,Bohn2017,Safronova2018}.  However, these advantages come at the cost of additional complexity, and  most molecules of interest are much more difficult to produce in useful form (e.g. a cold beam), than atoms. 

Cryogenic buffer gas cooling is general and has proved to be an essential starting point for many cold molecule experiments~\cite{Hutzler2012}, including precision measurements~\cite{Hutzler2011,Baron2014,ACME2018} and ultracold molecule production through direct laser cooling~\cite{Barry2011,Barry2014,Truppe2017Slow,Truppe2017SubDoppler,Anderegg2017,Collopy2018}.  Cryogenic buffer gas beam (CBGB) sources produce bright, slow molecular beams that are both translationally ($T$) and internally cold ($T_{int}$), typically with temperatures of $T\approx T_{int}\approx4$~K. In such sources, the molecular species of interest is introduced into a cryogenic cell containing a density-tuned, inert buffer gas (nearly always He or Ne). This is done via either a heated fill-line or laser ablation of a solid target. The resulting hot molecules, sometimes introduced at $T>1000$~K, are subsequently cooled by collisions with the buffer gas. Once thermalized, the molecular species is entrained within the cell in the buffer gas flow, and carried out of the cell through an aperture, forming a beam.  This cooling method is quite generic and can be applied to many species, from atoms to small bio-molecules, including highly reactive or refractory species~\cite{Hutzler2012}.

Current state-of-the-art molecular experiments that use CBGB sources~\cite{Anderegg2017,ACME2018,Collopy2018} are limited by the achievable molecular flux, and would benefit from generic methods to make more cold molecules.  In this manuscript, we report an order of magnitude increase in the molecular yield from a CBGB source by using laser light to excite a metal atom precursor. Specifically, we greatly increase the yield of polyatomic YbOH from our CBGB source by resonantly driving the $^1$S$_0 \rightarrow ^3$P$_1$ atomic Yb transition inside the buffer gas cell. The metastable ${}^3$P$_1$ state has a lifetime of $\tau\approx 871$~ns~\cite{Bowers1996}, long enough for the atoms to engage in reactive collisions before radiatively decaying, while also short enough to allow for rapid laser excitation. Our results establish laser-induced chemical enhancement via metastable excited states as a promising tool for significantly improving the production of cold molecules in CBGB sources, with significant implications for a broad range of precision measurement experiments.  

Reactive collisions involving excited species is a very active area within chemical physics. Depending on the species, promoting reactants to excited states can considerably modify the reaction dynamics and the product state distributions~\cite{Whitehead1988,Gonzalez-Urena1996,Telle2007}, with consequences for a wide range of fields, from astrophysics~\cite{Agndez2010,Zanchet2013,Fortenberry2011} to atmospheric chemistry~\cite{Cvetanovic1974,Wiesenfeld1982,Glowacki2012}. In many cases, the additional energy made available by electronic excitation of reactants can convert an endothermic reaction to an exothermic one. Additionally, the reaction mechanism on the excited potential energy surface can differ considerably from the mechanism for ground state reactants. As a result, excited states can access more pathways and transition states that yield the product of interest, as was seen in a recent study of Be$^+$ reactions~\cite{Yang2018}. 

In addition to modifying chemical yield, excited state chemistry has been used to study the collisional physics of atoms and molecules. In the case of atoms isoelectronic to Yb, such as Ca, Sr, Ba, and Hg, excitation of reactants to metastable states was used for molecular spectroscopy~\cite{Bernath1991,Bernath1997} and investigations of reactions in ovens or beams with gases such as SF$_6$~\cite{Gonzalez-Urena1995a,Gonzalez-Urena1996}, H$_2$~\cite{King1976b,Gonzalez-Urena1995a}, H$_2$O~\cite{Mestdagh1990,Davis1993,Oberlander1996c}, H$_2$O$_2$~\cite{Oberlander1991,Cheong1994}, alcohols~\cite{Mestdagh1990,Oberlander1996c,Gonzalez-Urena1996}, halogens~\cite{Gonzalez-Urena1995a,Gonzalez-Urena1996}, halogenated alkanes~\cite{King1976b,Solarz1979b,Gonzalez-Urena1996,Teule1999a,Husain2000a}, and hydrogen halides~\cite{Solarz1979b,Gonzalez-Urena1995a,Gonzalez-Urena1996,Teule1998c,DeCastro2000b}. %~\cite{King1976b,Solarz1979b,Oberlander1991,Davis1993,Cheong1994,Gonzalez-Urena1995a,Oberlander1996c,Teule1998c,Teule1999a,DeCastro2000b,Husain2000a}.
More recently, the ability to trap and cool species to ultracold temperatures has enabled research of reaction dynamics between excited ions, atoms, and molecules~\cite{Okada2003b,Yang2018,Puri2019,Mills2019}. 

Here, we characterize the excited state chemistry of a system of high interest for precision measurements~\cite{Kozyryev2017PolyEDM}. In particular, we study the dependence of enhanced YbOH yield on the properties of the laser light driving the atomic Yb transition, investigate the enhancement of various internal states, provide a simple model to interpret our observations, and perform computational studies of chemical reactions that produce YbOH.  Our computational results confirm the enhanced reactivity of the Yb($^3$P) state, and indicate the possibility of determining optimal reactants. Our experimental results also demonstrate the buffer gas environment effectively thermalizes the rotational and translational energies of the additional molecules produced via exothermic reactions with the excited atoms.

\section{Methods}

\begin{figure}[t]
    \centering
    \includegraphics{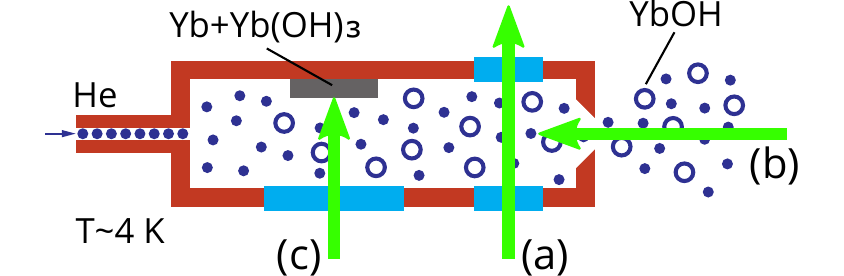}
    \caption{Depictions of the enhancement light geometries investigated. The enhancement light is denoted by the green arrow. (a) Transverse geometry: the enhancement light is introduced through a window $\sim25$~mm away from the ablation target and $\sim12$~mm away from the cell aperture.  (b) Longitudinal geometry: the enhancement light is introduced through the cell aperture.  (c) Collinear geometry: the enhancement light is sent through the ablation window, collinear with the YAG pulse.}
    \label{fig:geometries}
\end{figure}

Cold YbOH molecules are produced using cryogenic buffer gas cooling, which is reviewed in detail elsewhere~\cite{Hutzler2011,Hutzler2012}. Our source consists of a cryogenically cooled copper cell at $\sim 4$ K, depicted in Figure~\ref{fig:geometries}, which has an internal cylindrical bore with a diameter of 12.5 mm and a length of $\sim 100$~mm. Helium buffer gas enters the cell through a fill line at one end of the cell, and exits at the other end through an aperture 5~mm in diameter. The cell has windows that allow optical access for laser ablation and absorption spectroscopy. 

YbOH molecules are produced by laser ablation of a solid target with a pulsed nanosecond Nd:YAG laser at 532 nm. Unless stated otherwise, the data in this paper were taken with $\sim15$~mJ of energy at a repetition rate of 1-3 Hz. The data were obtained from targets of pressed Yb(OH)$_3$ powder in a stoichiometric mixture with Yb powder. The powders were mixed to have a 1:1 ratio of Yb and OH, ground using a mortar and pestle, passed through a 230 mesh sieve, mixed with 4\% PEG8000 binder by weight, and pressed in a die with 8~mm diameter at a pressure of 10~MPa for $\sim15$ minutes. The behavior of the laser-induced enhancement was found to be similar for variety of other targets with slightly different compositions. From such targets, a single ablation shot typically produces $\sim 10^{14}$ thermalized Yb atoms, orders of magnitude more than typical yields of molecular YbOH.

To study molecular production, we use a 577~nm laser to perform absorption spectroscopy on the $^{Q}Q_{11}(2)$ line of the $\Tilde{X} ^2 \Sigma^{+} (000) \rightarrow \Tilde{A} ^2 \Pi_{1/2} (000)$ transition in $^{174}$YbOH. Here, $(v_1 v_2 v_3)$ denote the vibrational quanta in the Yb-O stretch, O bend, and O-H stretch, respectively. For $^{174}$YbOH transitions, we use the labeling scheme described in Ref.~\cite{Lim2017} and references therein. The laser light is produced by doubling a 1154~nm ECDL using a PPLN waveguide. Absorption of the probe was used to determine the number density of molecules both inside the cell and immediately in front of the cell aperture. Unless stated otherwise, Yb refers to $^{174}$Yb for both atomic Yb and YbOH. 

\begin{figure}[t]
    \centering
    \includegraphics{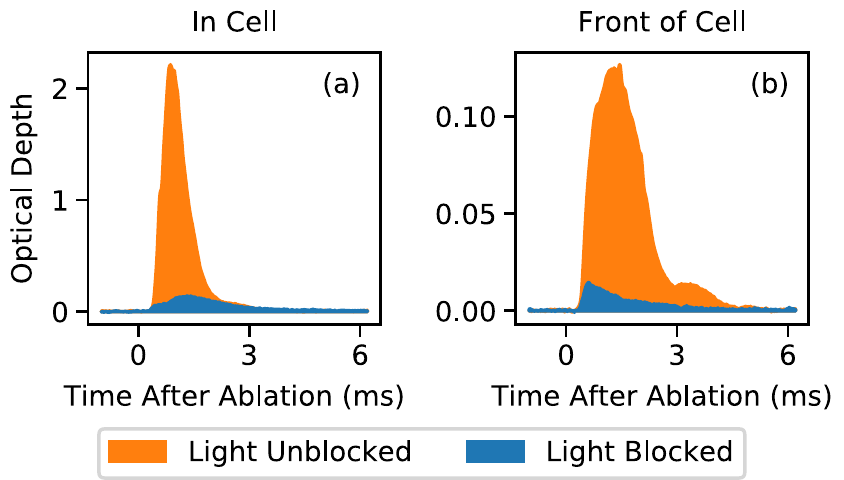}
    \caption{\label{fig:en_OD} Absorption spectroscopy of YbOH density in the  $N=2$, $\tilde{X}^2 \Sigma^{+} (000)$ state, both in-cell and front-of-cell. (a): In-cell un-enhanced yield of $4\times10^{10}$ molecules, enhanced yield of $3\times10^{11}$ molecules. (b): Front-of-cell un-enhanced yield of $7\times10^9$ molecules, enhanced yield of $8\times10^{10}$ molecules.}
\end{figure}

To enhance the production of molecules, we excite the 556~nm $^{1}$S$_0 \rightarrow {}^{3}$P$_1$ transition in atomic Yb.  The light is derived by sum-frequency generation of a CW Ti:Saph with a 1550~nm fiber laser, and has a linewidth of $<50$~kHz~\footnote{Sirah Mattise Ti:Saph and NKT ADJUSTIK+BOOSTIK combined in a Sirah MixTrain.}.  The light is pulsed on and off with a combination of an acousto-optical modulator (AOM) and mechanical shutter, allowing us to study the effect of the excitation timing relative to the ablation pulse.  The mechanical shutter passes the light into the cell $\sim 4$~ms before the ablation pulse, and blocks the light again $\sim 8$~ms after the ablation, in order to keep the cell from being heated unnecessarily.  The AOM is used in conjunction with the shutter to perform more precise measurements of the effects of pulse timing.  The shutter stays closed for every other molecule pulse, in order to normalize against drifts in molecular yield as the ablation spot degrades.

\section{Results}

To determine the molecular yield inside the cell and the molecular flux leaving the cell, we integrate the optical depth (OD) over the duration of the resulting $\sim$ ms long molecule pulse. Figure \ref{fig:en_OD} shows a representative absorption signal from a single ablation shot, both with and without the enhancement light present. We compute the enhancement factor, or fractional increase in the number of molecules, by taking the ratio of the integrated OD with and without the enhancement light. Since the probe light is always fixed at the same molecule transition, common factors such as cross section divide out, making the OD ratio directly sensitive to changes in molecule number density. Typically, in-cell YbOH population in the $N=2$, $\tilde{X}^2\Sigma^+(000)$ state was enhanced from $\sim 10^{10}$ to $\sim 10^{11}$, with front-of-cell numbers similarly enhanced, from $\sim 10^9$ to $\sim 10^{10}$ molecules. The enhancement factor depends on a number of parameters, such as laser power, detuning, timing, and geometry, which we will now discuss. 

We investigated three geometries for introducing the enhancement light into the cell, indicated in Figure~\ref{fig:geometries}. The 556 nm light was typically collimated to a beam diameter of $\sim2.5$~mm. The largest enhancement signals were observed when the light was sent through the window in the cell used for absorption spectroscopy, shown in Figure~\ref{fig:geometries}(a). For a given target, the enhancement effect was repeatable for different ablation sites. For the second geometry, shown in Figure~\ref{fig:geometries}(b), the light entered the cell longitudinally through the circular, 5~mm diameter cell aperture. The resulting enhancement magnitude was reduced by a factor of $\sim 2$, with the effect somewhat independent of the ablation site. In the final geometry involved the enhancement light overlapped with the path of the ablation laser, shown in Figure~\ref{fig:geometries}(c). When compared to the aforementioned geometries, this co-linear geometry resulted in smaller and less consistent enhancement. Unless stated otherwise, the data in this paper are from the first geometry, with the enhancement light sent through the spectroscopy window.

\begin{figure}[t]
    \centering
    \includegraphics{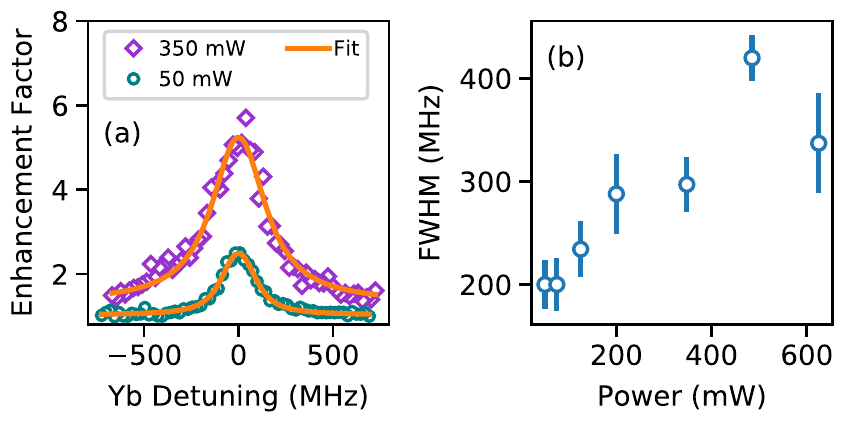}
    \caption{\label{fig:en_chirps} Enhancement line shapes, data taken with the longitudinal geometry. Left: Frequency scans and Voigt fits, demonstrating the variation of YbOH enhancement with detuning of the Yb laser at different powers. Right: Full widths at half maximum for the enhancement line shape as a function of the power sent into the cell.  The Doppler width for the Yb atomic transition averaged over the entire ablation pulse is $\sim 150$~MHz.}
\end{figure}

To characterize the frequency dependence of the enhancement, we scanned the enhancement laser frequency across the atomic Yb line while monitoring the YbOH yield with a resonant absorption probe. The resulting enhancement magnitude for such scans at varied powers is shown in  Figure~\ref{fig:en_chirps}, demonstrating the resonant nature of the enhancement. Since we apply sufficient laser power to power broaden the transition by an amount comparable to the Doppler broadening, we successfully fit the shape to a Voigt distribution. The extracted full-widths-at-half-maximum (FWHM), obtained from frequency scans in the longitudinal geometry, are plotted against enhancement power in the inset of Figure~\ref{fig:en_chirps}.   

The observed enhancement widths indicate a broader reactant Yb frequency distribution than that expected from Doppler broadening at $\sim 4$ K and power broadening from $\sim 200$~mW of resonant light. A similarly broad distribution is observed from low intensity scans of the atomic line shape alone. At ablation energies of $\sim 15$~mJ, the first $<1$~ms of the Yb absorption trace contribute to significant broadening, indicating the presence of an early, athermal Yb population~\cite{Skoff2011}. The remaining population present after 1 ms are consistent with a Doppler broadening at $T\sim4$~K. Because the enhancement light can excite this early athermal Yb population, we expect the atoms to react, providing the additional broadening we observe in the enhancement line shape.  A typical value for the FWHM of a Doppler-broadened Yb atomic absorption line (in the limit of low saturation parameter) is $\sim70$~MHz if the athermal component is excluded, and $\sim150$~MHz if it is included.

\begin{figure}[t]
    \centering
    \includegraphics{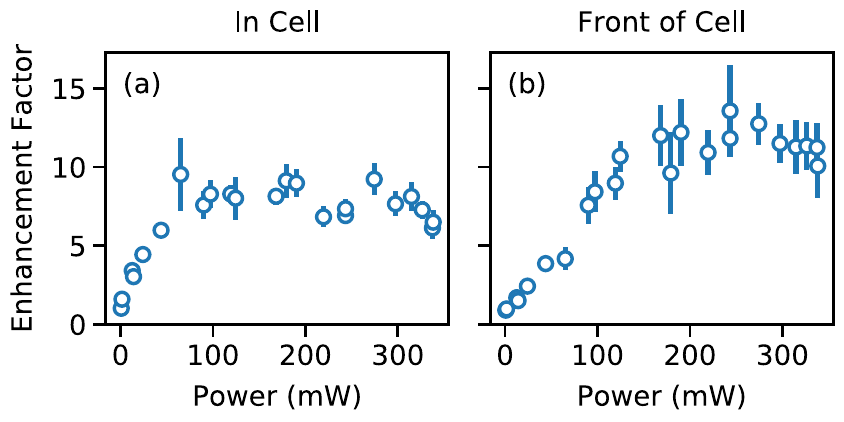}
    \caption{\label{fig:en_power} Enhancement magnitude, calculated as a ratio of optical depths, plotted against the laser power sent into the cell. The laser beam was collimated to a $\sim2.5$~mm diameter. Error bars represent standard deviations of results, as opposed to standard error, in order to show the typical fluctuations over different shots and ablation spots. (a): In cell enhancement, (b): Front of cell enhancement.}
\end{figure}

The enhancement factor has a nonlinear dependence on the power of the enhancement light. This relationship is illustrated in Figure \ref{fig:en_power}, showing the transition of the enhancement magnitude from linear behavior at low powers to saturation at high power. The crossover typically occurs between 100 and 300 mW for a $\sim2.5$~mm beam, corresponding to an intensity range of $\sim10$ W/cm$^2$.  Such behavior is indicative of driving an optical resonance, and supports a simple model where the enhancement magnitude is proportional to the steady state excited Yb population. Notice that this cross-over intensity is considerably higher than the saturation intensity of the transition (0.14~mW/cm$^2$), which is due to the fact that the transition is Doppler broadened~\cite{Budker2008}. We expect the effect to saturate when the power broadening is comparable to the Doppler broadening~\cite{Budker2008}. The power broadened radiative width is $\gamma_{tot} \approx \gamma_{rad}\sqrt{s}$, where $\gamma_{rad}\approx180$~kHz is the natural width and $s$ is the saturation parameter. The broadened width becomes comparable to the Doppler width $\delta_{D}\approx70$~MHz when $s\approx(\delta_{D}/\gamma_{rad})^2\approx10^5$, or $I\approx10$~W/cm$^2$, consistent with our measurements.

By using an AOM switch to pulse the atomic transition light for sub-ms duration, we determined the majority of the enhancement occurs in the first few ms after ablation, corresponding to the duration when the cell is filled with atomic Yb. Notably, the enhancement is largest $\sim1$~ms after the ablation, after the hot atoms have thermalized with the buffer gas. This observation, combined with the effect of geometry on enhancement, provides evidence that the enhancement occurs throughout the cell, rather than immediately in the region of the ablation plume. 

\begin{figure*}[t]
    \centering
    \subfloat{
    \includegraphics{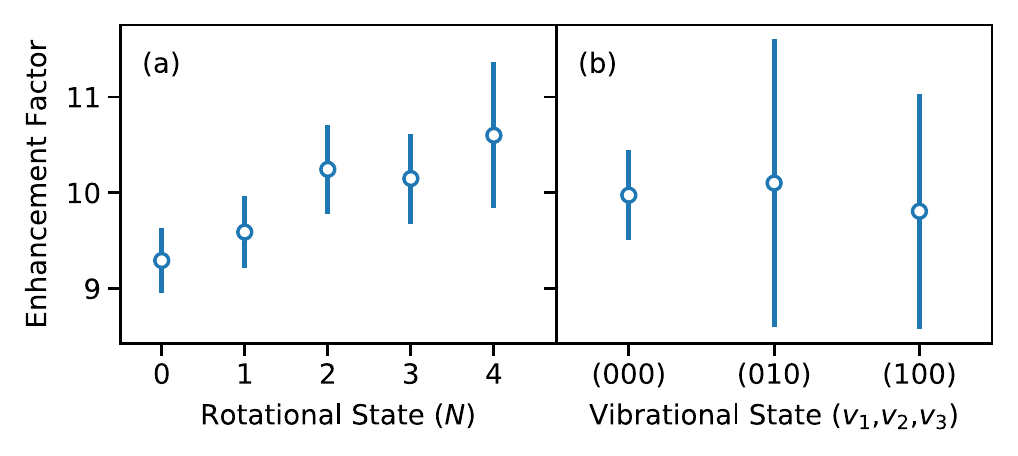}
    \label{fig:en_rot_vib}
    }
    \subfloat{
    \includegraphics{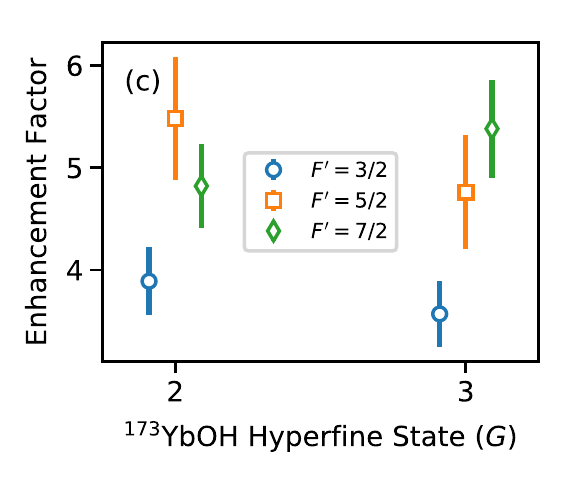}
    \label{fig:en_173}
    }
    \captionsetup{singlelinecheck=off}
    \caption{Enhancement of rotational and vibrational states in $^{174}$YbOH, and hypefine states of $^{173}$YbOH. Error bars represent the standard deviation of measured enhancement factors. (a), (b): Enhancement of $^{174}$YbOH as a function of ground state rotational level (a) and ground state vibrational level (b). The rotational population was probed using $^QQ_{11}(N)$ lines. $(v_1 v_2 v_3)$ denote the vibrational quanta in the Yb-O stretch, O bend, and O-H stretch respectively. The $(000)$ data point is an average of the $N=0$ through $N=4$ rotational enhancements. The excited vibrational population was probed with diagonal transitions to the $\tilde{A}$ state with $\Delta v_{1,2}=0$. (c): Enhancement of the molecular hyperfine levels in the odd $^{173}$YbOH isotopologue, resulting from driving $F=5/2\rightarrow F^\prime$ hyperfine transitions in atomic $^{173}$Yb. The molecular quantum number $G$ results from coupling of $S$ to $I_{Yb}$, $G=S+I_{Yb}$.}
    \label{fig:enhancement vs N,vib,isotope}
\end{figure*}

The enhancement magnitude was not found to have any significant dependence on He flow into cell, which was varied from 1 to 10 standard cubic centimeters per minute (SCCM), equivalent to varying the stagnation He density in the cell from $6\times10^{14}$ to $6\times10^{15} \mathrm{cm}^{-3}$~\cite{Hutzler2012}. The enhancement magnitude was also unaffected by the ablation energy used in the ablation pulse, which was varied from 5 to 25 mJ/pulse. In fact, for low ablation energies, YbOH was observed only with the aid of enhancement, as long as the ablation energy was above the threshold necessary to produce atomic Yb. This is encouraging for laser-cooling experiments, where lower energy ablation is useful for producing slow beams of molecules~\cite{Anderegg2017}. 

We also investigated the effect of the enhancement light on the population of YbOH in different internal states. Since the energy scales of the chemical reactions involved are on the order of several thousand cm$^{-1}$, much larger than those of molecular vibration (hundreds of cm$^{-1}$) or rotation (tens of cm$^{-1}$), we expect that the molecules created by chemical reactions will populate many rotational and vibrational states after decaying to the ground electronic state. These distributions have been studied in excited state reactions producing molecules containing Ca and Sr, and they support the expectation that the released energy is distributed among the internal modes~\cite{Cheong1994,Oberlander1996c}. 

Because rotational state-changing cross sections between molecules and helium are comparable to elastic collision cross sections~\cite{Hutzler2012}, we expect this broad rotational distribution to rapidly thermalize in the buffer gas cell. By measuring the enhancement on $^QQ_{11}(N)$ transitions that address different rotational levels in the ground vibronic state, we indeed observe such rotational thermalization, as shown in Figure~\ref{fig:enhancement vs N,vib,isotope}(a). Each rotational transition demonstrates approximately the same enhancement, indicating that the rotational distribution is essentially unchanged by the increased chemical production.

Since buffer gas collisions are also effective at thermalizing translational degrees of freedom, we expect the enhanced and un-enhanced molecule beams to have similar velocity properties. We verified this by monitoring the transverse velocity distribution of YbOH exiting the cell using an absorption probe in front of the cell aperture. The width of the resulting line shapes did not exhibit a measurable difference with and without the enhancement. Similarly, we monitored Doppler shifted fluorescence of the molecular beam $\sim60$~cm downstream, after a series of collimating apertures, and found the both the mean and width of the forward velocity distribution were unaffected by the enhanced molecular yield. 

Conversely, vibration-quenching cross sections are typically smaller than those for other degrees of freedom, resulting in observations of non-thermal vibrational distributions in CBGB sources~\cite{Hutzler2012, Kozyryev2015,Bu2017}. The efficiency of vibrational thermalization can vary for different molecular species, as well as for different modes of the same molecule~\cite{Kozyryev2015}. In our source, we observe non-thermal vibrational distributions, probed by absorption of diagonal transitions ($\Delta v=0$) from the (100) and (010) vibrational states in $\tilde{X}$ to the same vibrational state in $\tilde{A}$, located at $17345.09$~cm$^{-1}$ for the O bend, and $17378.58$~cm$^{-1}$ for the Yb-O stretch~\footnote{T. C. Steimle, Private Communications (2019).}. The populations we observe in these vibrational states, $\sim1$~ms after ablation, correspond roughly to temperatures of $T_{v_1} \approx 280$~K and $T_{v_2} \approx 110$~K. Our observations of more effective thermalization for lower-lying vibrational states is in agreement with a recent study of SrOH thermalization in a buffer gas cell~\cite{Kozyryev2015}.

The vibrationally excited molecule population in the cell was also significantly enhanced by laser excitation of Yb. In Figure \ref{fig:enhancement vs N,vib,isotope}(b) we compare the enhancement for the (000), (100), and (010) vibrational levels of the $\tilde{X}$ ground electronic state. We find the enhancement factor to be consistent across these vibrational states, indicating buffer gas collisions do not efficiently quench the vibrational states populated by the excited state chemistry. Enhanced yield in vibrational states can be desirable, as excited vibrational levels may have little population in a typical beam source, but are required for laser cooling, spectroscopy, and precision measurements~\cite{Kozyryev2017PolyEDM}. Furthermore, these vibrational populations can be easily ``re-pumped'' back into the ground state, e.g. using the same lasers that would already be available for laser cooling, resulting in further increases to beam brightness.

Finally, we characterize the enhancement in the $^{173}$YbOH isotopologue, which has high sensitivity to the symmetry-violating nuclear Magnetic Quadrupole Moment (NMQM)\cite{Kozyryev2017PolyEDM,Maison2019}, by investigating the enhancement of different $^{173}$YbOH hyperfine states when driving hyperfine transitions in atomic $^{173}$Yb $(I=5/2)$. The results are shown in Figure \ref{fig:enhancement vs N,vib,isotope}(c). We separately drive each of the three ${}^1$S$_0 \rightarrow {}^3$P$_1$ hyperfine transitions in $^{173}$Yb ($F=5/2\rightarrow F'=3/2,5/2,7/2$) and monitor the enhancement in either the $G=2$ or $G=3$ hyperfine state of $^{173}$YbOH. Here, analogous to the case of $^{173}$YbF~\cite{Wang2019}, the coupled angular momentum $G=S+I_{\textrm{Yb}}$ results from the strong electric quadrupole interaction between the Yb-centered electron, with spin $S=1/2$, and the non-spherical Yb nucleus, with spin $I_{\textrm{Yb}}=5/2$. The molecule population was probed via absorption spectroscopy on the $^OP_{12}(2)$ and $^OP_{13}(2)$ lines of the $\Tilde{X}\rightarrow\Tilde{A}$ transition, where we label the transitions using the convention from Ref.~\cite{Wang2019}. The enhancement in the $G=2$ and $G=3$ states is equivalent for each driven $^{173}$Yb hyperfine transition, which is expected in a thermalized ensemble. While thermalization should also result in enhancement independent of the excited hyperfine $F'$ state driven in $^{173}$Yb, we find smaller enhancement for $F'=3/2$ compared to $F'=5/2$ and $F'=7/2$.  We attribute this to overlap of the $^{173}$Yb$(F=5/2 \rightarrow F^{\prime}=3/2)$ transition with the $^{171}$Yb$(F=1/2 \rightarrow F^{\prime}=3/2)$ transition, which differ by $\sim3$~MHz~\cite{Pandey2009}, much less than the Doppler broadening in the cell. This overlap can explain lower enhancement rates, as the production of $^{171}$YbOH will deplete the available population of other reactants. 

\section{Chemistry}

To elucidate the chemistry behind our experimentally observed enhancement of YbOH, we performed calculations of electronic structure and molecular dynamics. Specifically, we considered the cases of Yb reacting with H$_2$O and H$_2$O$_2$, two reactants likely produced during laser ablation of solid targets containing Yb(OH)3.

Geometry optimizations for the individual reactants, intermediate complexes, and reaction products were performed using non-relativistic Density Functional Theory (DFT) with the UCAM-B3LYP functional~\cite{YANAI200451}. For the H$_2$O and H$_2$O$_2$ molecules, we use the aug-cc-pVTZ basis set~\cite{dunning:89}, while for the Yb($^1$S) and Yb($^3$P) states, we chose the Stuttgart Effective Core Potential ECP28MWB~\cite{Dolg:1989}, combined with the def2-QZVPP basis set~\cite{Gulde:2012,BSexch:2019}. 

\begin{figure*}[t]
    \centering
    \includegraphics[width=0.8\textwidth]{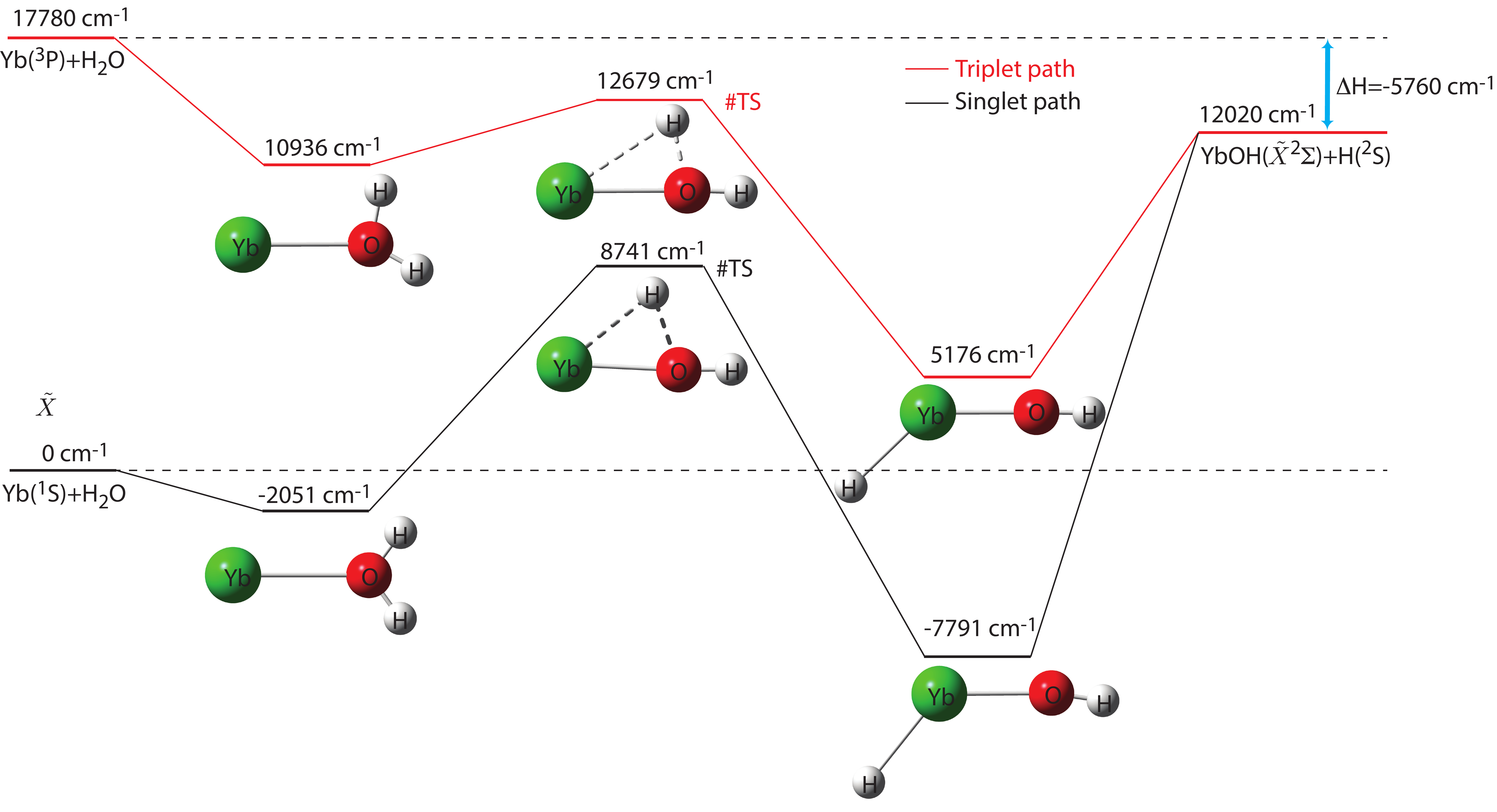}
    \caption{\label{fig:ybh2odiag} Energy profile for the Yb($^1$S)+H$_2$O and Yb($^3$P)+H$_2$O reactions leading to YbOH($\tilde{X}^2\Sigma^+$)+H($^2$S) products calculated at the DFT/UCAM-B3LYP level of theory~\cite{YANAI200451}. The molecular models represent the system geometries at critical points, and were drawn in the Gauss View 5 program~\cite{gv5}. The Yb, O, and H atoms are represented by green, red and white spheres, respectively. Solid and dashed lines connecting the atoms correspond to  $\sigma$ bonds and temporary connections the transition states, respectively.}
\end{figure*}

The intermediate complexes formed along the reaction paths of Yb($^1$S)+H$_2$O/H$_2$O$_2$ and Yb($^3$P)+H$_2$O/H$_2$O$_2$ were optimized to their minima or saddle points, corresponding to transition states. The transition states were found by the Synchronous Transit-Guided Quasi-Newton method~\cite{QST2:1993}, implemented in the Gaussian09 program~\cite{Gaussian09_RevE}. Frequency calculations were also performed to ensure the geometry is optimized to the minimum or saddle point (indicated by a single imaginary frequency).

We calculate an {\em ab initio} value of $E/hc=17780$ cm$^{-1}$ for the transition energy between the $^1$S and $^3$P states of Yb. The quality of our DFT calculations can be characterized by comparison of our {\em ab initio} value with the experimental value of 17992 cm$^{-1}$ for the $^1$S$_0\rightarrow^3$P$_1$ transition~\cite{Sansonetti_Handbook:2005}.

\subsection{Yb+H$_2$O}

\begin{figure*}[t]
    \centering
    \includegraphics[width=0.8\textwidth]{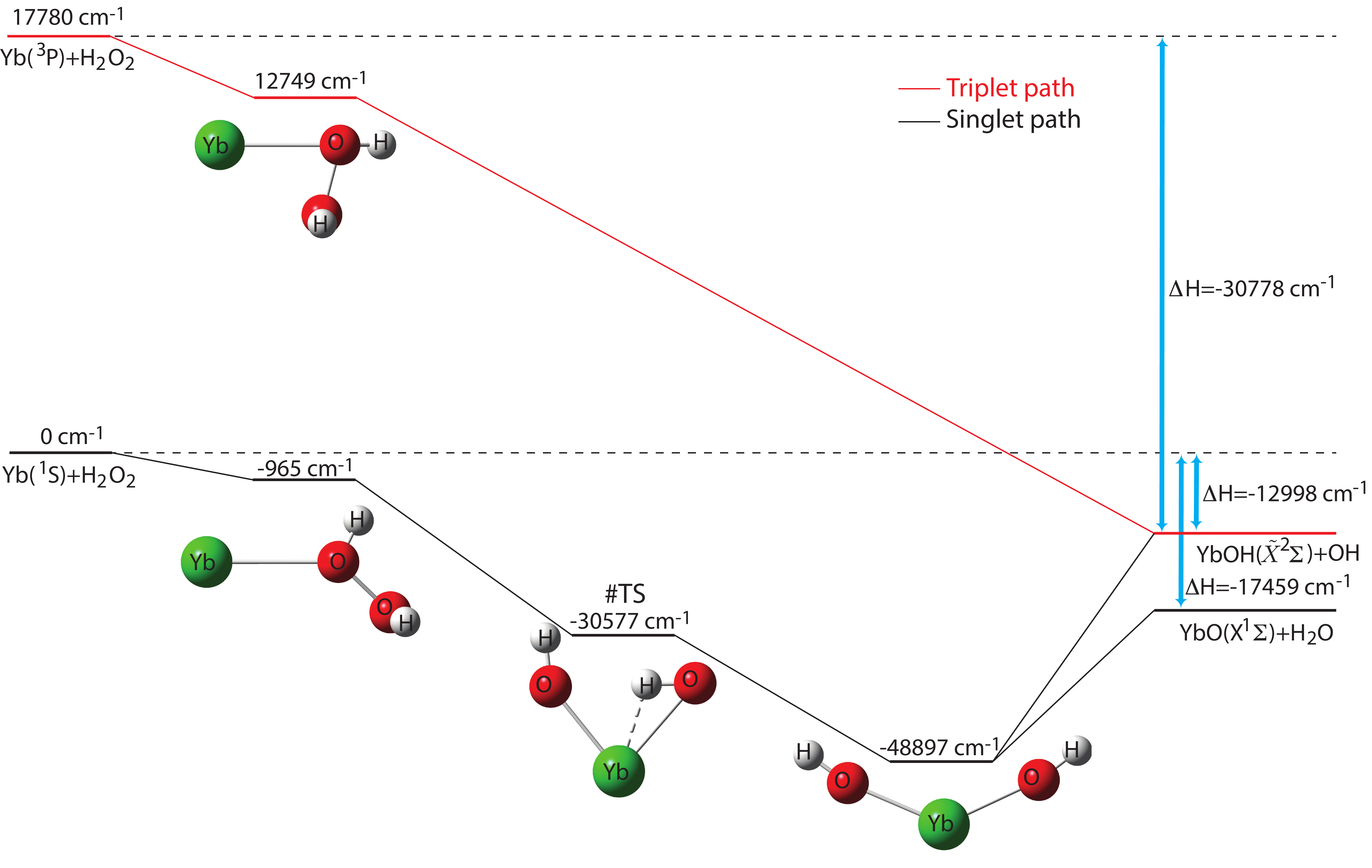}
    \caption{\label{fig:ybh2o2diag} Energies and molecular geometries at critical points for the Yb($^1$S)+H$_2$O$_2$ collision and Yb($^3$P)+H$_2$O$_2$ reaction leading to either YbO($X^1\Sigma^+$)+OH($X^2\Pi$)  and YbOH($\tilde{X}^2\Sigma^+$)+H  calculated by DFT and drawn with Gauss View 5 ~\cite{gv5}. The Yb, O, and H atoms are represented by green, red and white spheres, respectively.  Solid and dashed lines connecting the atoms correspond to  $\sigma$ bonds and temporary connections the transition states, respectively.}
\end{figure*}

Figure~\ref{fig:ybh2odiag} shows our correlation diagram, i.e. the critical points on high-dimensional potential energy surfaces, for the Yb($^1$S)+H$_2$O and Yb($^3$P)+H$_2$O reactions. Firstly, our calculations demonstrate that the reaction between a ground state Yb($^1$S) atom and an H$_2$O molecule is endothermic, requiring $E/hc=12020$ cm$^{-1}$ of relative kinetic energy to proceed and form the product YbOH($\tilde{X}^2\Sigma^+$)+H($^2$S). Secondly, this singlet potential energy surface has a transition state, or saddle point, that lies at 8741 cm$^{-1}$ above the entrance channel. It separates a local minimum corresponding to a symmetric-top molecule, where none of the bonds within H$_2$O are significantly affected by Yb, and the global minimum, where one of the hydrogen atoms has broken from the water molecule and the Yb atom is inserted.  

In contrast, the collision between the excited Yb($^3$P) state and H$_2$O is exothermic by 5760 cm$^{-1}$. Moreover, the corresponding triplet potential energy surface has a transition state that lies below its asymptotic channel energies. Such a submerged reaction barrier is indicative of large reaction rates. For both singlet and triplet channels, the product YbO+H$_2$ is energetically inaccessible, lying 27233~cm$^{-1}$ above the Yb($^1$S)+H$_2$O reaction channel.

\subsection{Yb+H$_2$O$_2$ }

Figure~\ref{fig:ybh2o2diag} shows our correlation diagram for the relevant spin singlet and triplet Yb+H$_2$O$_2$ reactions. For both Yb($^1$S)+H$_2$O$_2$ and Yb($^3$P)+H$_2$O$_2$ collisions, the product molecules have a lower electronic energy than the initial reactants. In fact, both YbO($X^1\Sigma^+$) + H$_2$O and YbOH($\tilde{X}^2\Sigma^+$) + OH($X^2\Pi$) products are energetically accessible, in contrast to the reaction with H$_2$O.  

The relative kinetic energy of the product molecules is significantly larger than that for the product in the Yb($^3$P)+H$_2$O reaction. The transition state on the spin singlet potential surface is submerged, and its global minimum corresponds to a deeply-bound (OH)-Yb-(OH) molecule. We thus expect strong reactivity along this pathway. Finally, we did not find a transition state on the spin triplet surface, and spin conservation implies that only YbOH($\tilde{X}^2\Sigma^+$) + OH($X^2\Pi$) can be formed. 

\subsection{Molecular Dynamics Simulations} 

To further investigate the reactivity of atomic Yb colliding with H$_2$O and H$_2$O$_2$, we performed Born-Oppenheimer Molecular Dynamics (BOMD)~\cite{HELGAKER1990145} calculations on both the singlet and triplet potential energy surfaces, as implemented in Gaussian09~\cite{Gaussian09_RevE}, with the basis sets used above. The initial condition of the dynamics is set to a rotational temperature of $4$~K, corresponding to the temperature of CBGB source in our experiment. The Yb atom is initialized $\sim4$~\AA~away from the O atom in H$_2$O or H$_2$O$_2$. The molecular dynamics are performed with the statistical NVE ensemble and single trajectory. Movies of the classical trajectories for the motion of Yb($^1$S) and Yb($^3$P) colliding with H$_2$O and H$_2$O$_2$ are presented in Supplementary Material.

The simulations show the Yb($^1$S)+H$_2$O system forms a YbH$_2$O complex, without reacting and producing YbOH product molecules. For the Yb($^3$P)+H$_2$O collision, the HYbOH intermediate forms immediately, after which the hydrogen atom attached to the Yb quickly flies away, leaving the YbOH product. Our simulations are in agreement with previous studies of Ca($^3$P), Sr($^3$P), and Ba($^1$D) reacting with H$_2$O and preferentially forming metal hydroxides~\cite{Davis1993,Oberlander1996c}.  

In the case of Yb($^3$P)+H$_2$O$_2$, the dynamics show that the spin singlet reaction produces the YbO, and the spin triplet forms YbOH. A similar dependence of MO/MOH (M = metal atom) product on the reactant electronic state was previously observed in reactions of Ca and Sr with H$_2$O$_2$~\cite{Oberlander1991,Cheong1994}. The Yb($^1$S)+H$_2$O$_2$ reaction occurs over 420~fs, slower than the 120~fs and 180~fs reaction times for Yb($^3$P) reacting with H$_2$O$_2$ and H$_2$O, respectively. The difference in reaction times may indicate a difference in reaction mechanism between the two atomic states. Our simulations confirm the increased reactivity of Yb($^3$P), even when the products are also energetically accessible for Yb($^1$S).

\section{Discussion}

By driving an electronic transition from Yb($^1$S$_0$) to Yb($^3$P$_1$), we have demonstrated significantly improved yield of molecular YbOH from a CBGB source. The resonant nature of the effect, as well as saturation at high power, confirms that the excited atomic population is responsible for the observed enhancement. By performing computational studies, we are able to provide insight into the reaction channels made possible by excited Yb($^3$P) atoms. Furthermore, we found that the cryogenic buffer gas environment is well suited to cooling the products from the resulting exothermic reactions. Buffer gas collisions effectively thermalized the translational and rotational energies of the resulting product molecules, while still maintaining an athermal vibrational population, which is useful for many applications. 

Our approach suggests a number of new directions for both further improvements to molecular yield in future experiments and continued studies of cold chemical reactions. From our studies of geometry and timing, the enhancement occurs throughout the cell and over the entire duration of the molecular pulse, suggesting an optimal arrangement where the cell is evenly illuminated with resonant light. Although we used only a solid precursor in the studies presented here, another approach is to use reactant gases flowed into the buffer gas cell via a capillary~\cite{Patterson2007,Anderegg2017,Truppe2017Slow}. These molecular precursors react with ablated metal, providing a way to tune the reactant species. While the enhancement we report here is a compound effect, possibly involving several different reactants formed in ablation, our calculations suggest the possibility of finding the optimal reactant and optimal excited states for both the atom and molecule. In addition, enhancing reaction rates would allow for reduction of ablation energy without also compromising molecular flux. 

While we have restricted our measurements to YbOH, it is likely that this method can be used to enhance CBGB production of many interesting species, both diatomic and polyatomic.  The chemical similarity of Yb with alkaline earth atoms, plus the success of excited state chemical reactions producing a variety of Ca-, Sr-, and Ba- molecules with numerous ligands~\cite{Davis1993,Oberlander1996c,Bernath1997}, suggests that CBGBs of alkaline-earth atoms with monovalent and ionic bonds (conveniently, those which can be generically laser cooled~\cite{Isaev2016Poly}) could benefit from this approach.  Note, however, that the power requirements become higher for lighter species, since the radiative width of the metastable states arises from spin-orbit coupling, which is larger in heavier species~\cite{Santra2004}.  Nonetheless, resonant excitation of the metal precursor could be especially helpful for experiments with rare isotopes where efficiency is critical, such as $^{225}$Ra, which is a component of molecules with extremely high sensitivity to physics BSM~\cite{Isaev2010,Isaev2017RaOH}, or $^{26}$Al, which is of astrophysical relevance~\cite{Kaminski2018}.  While we have mostly focused on alkaline-earth or similar metals, CBGBs of other molecules of experimental importance, such as ThO~\cite{ACME2018}, may also benefit from this approach by exciting the metal~\cite{Au2014} or oxygen~\cite{Gonzalez-Urena1995a} produced in the ablation to a reactive, metastable state.

In addition to increasing CBGB yield, chemical enhancement can also serve as a resource for spectroscopy of dynamics inside the buffer gas cell. The dependence of the molecular yield on the application of enhancement light at a specific time and place can help study the distribution of the reactive dynamics in the cell. When compounded with probes monitoring the flux exiting the cell, or monitoring fluorescence downstream, this allows for study of beam properties, conditioned on where or when the molecules were produced. The ability to perform such spectroscopy could aid in understanding and optimizing buffer gas cell geometries. 

Our enhancement method could also be used to disentangle complex spectroscopic data by comparing enhanced and normal spectral features, taking into account the enhancement dependence on the excited atomic state, as well as the molecular vibrational, rotational, and hyperfine state.  For example, the spectra of hypermetallic species~\cite{ORourke2019} could be uniquely distinguished from other molecules by their dependence on the chemical enhancement of the individual metal centers.  Additionally, because the molecules resulting from enhancement can possibly populate vibrational states non-thermally, yet still yield translationally cold beams, this technique will be useful for studying transitions out of excited vibrational modes. The increased vibrational population is also favorable for precision measurements~\cite{Kozyryev2017PolyEDM}, spectroscopy, and studies of vibration-quenching collisions in cryogenic environments~\cite{Kozyryev2015}.

Finally, for precision measurements relying on CBGBs, increased molecular flux directly translates to increased sensitivity to new, symmetry-violating physics beyond the Standard Model. Specifically, the enhancement we demonstrate for both the $^{174}$YbOH and $^{173}$YbOH isotopologues are directly applicable to experiments sensitive to new physics in both the leptonic and hadronic sectors~\cite{Kozyryev2017PolyEDM, Gaul2018, Maison2019, Denis2019, Prasannaa2019}. 

Recently, Ref.~\cite{Augenbraun2019} demonstrated Doppler and Sisyphus cooling of $^{174}$YbOH, an isotopologue with high sensitivity to the electron's electric dipole moment (eEDM). By combining the $^{174}$YbOH flux obtained from the enhanced source described here with laser slowing, magneto-optical trap loading, and transfer to an optical dipole trap (ODT), we estimate that an eEDM experiment with $\sim 10^5$ molecules in an ODT is feasible~\footnote{This estimate is based on previously reported efficiencies for slowing and trapping diatomic molecules from CBGBs~\cite{Anderegg2017,Truppe2017Slow,Cheuk2018}}. This would lead to an eEDM sensitivity surpassing the current limit of $d_e<1.1\times10^{-29}$~e~cm~\cite{ACME2018}. This sensitivity could be further increased by additional technical improvements, such as beam focusing~\cite{Kaenders1995}, transverse confinement~\cite{DeMille2013}, and ``few photon" slowing techniques~\cite{Lu2014,Fitch2016}. 

\section*{Acknowledgements}

We thank the PolyEDM collaboration for insightful discussions, especially  Amar Vutha, Tim Steimle, John Doyle, and the Doyle Group at Harvard.  We also thank Tim Steimle for providing the lines used for spectroscopy of YbOH, Avikar Periwal for his assistance fabricating the CBGB source, Yi Zeng for assistance with the lasers, and Elizabeth West, Xiaomei Zeng, and Katherine Faber for assistance with fabricating the ablation targets.  AJ thanks Ashay Patel for helpful discussions. 

Work at Caltech is supported by the Heising-Simons Foundation, the NSF CAREER Award \#PHY-1847550, and NIST Precision Measurement Grant
\#60NANB18D253. 

Work at Temple University is supported by the Army Research Office Grant \#W911NF-17-1-0563, the U.S. Air Force Office of Scientific Research Grant \#FA9550-14-1-0321 and the NSF Grant \#PHY-1908634.

\bibliographystyle{apsrev4-2}
\bibliography{references}

\end{document}